\documentclass[11pt,a4paper]{article}
\usepackage{jcappub}
\usepackage[utf8]{inputenc}
\pdfoutput=1

\usepackage{graphicx}%
\usepackage{multirow}%
\usepackage{amsmath,amssymb,amsfonts}%
\usepackage{amsthm}%
\usepackage{mathrsfs}%
\usepackage[title]{appendix}%
\usepackage{xcolor}%
\usepackage{textcomp}%
\usepackage{manyfoot}%
\usepackage{booktabs}%
\usepackage{algorithm}%
\usepackage{algorithmicx}%
\usepackage{algpseudocode}%
\usepackage{listings}%
\usepackage{comment}
\usepackage[normalem]{ulem}
\usepackage[utf8]{inputenc}
\usepackage{lmodern}         
\usepackage{fontawesome}

\newcommand{\be}{\begin{equation}}
\newcommand{\ee}{\end{equation}}
\newcommand{\beq}{\begin{equation*}}
\newcommand{\eeq}{\end{equation*}}
\newcommand{\bea}{\begin{eqnarray}}
\newcommand{\eea}{\end{eqnarray}}

\newcommand{\bn}{{\mathbf n}}
\newcommand{\bk}{{\mathbf k}}

\newcommand{\bv}{{\mathbf v}}

\newcommand{\diff}{\mathrm{d}}
\newcommand{\HH}{{\cal H}}

\newcommand{\De}{\Delta}

\newcommand{\Omm}{\Omega_{{\rm m}}}

\newcommand{\al}{\alpha}

\newcommand{\de}{\delta}

\newcommand{\La}{\Lambda}
\newcommand{\si}{\sigma}

\newcommand{\Om}{\Omega}

\newcommand{\cd}{\cdot}

\definecolor{dgreen}{rgb}{0,0.6,0.0}

\definecolor{dblue}{rgb}{0,0,0.7}

\begin{document}

\title{Anchors no more:\\ Using peculiar velocities to constrain $H_0$ and the primordial Universe without calibrators}

\author[a,b,*]{Davide Piras,}
\author[b,*]{Francesco Sorrenti,}
\author[b]{Ruth Durrer,}
\author[b]{Martin Kunz}


\affiliation[a]{Centre Universitaire d’Informatique, Université de Genève,\\7 route de Drize, 1227 Genève, Switzerland}
\affiliation[b]{Département de Physique Théorique, Université de Genève,\\24 quai Ernest Ansermet, 1211 Genève 4, Switzerland}
\affiliation[*]{Equal contribution}

\emailAdd{davide.piras@unige.ch}
\emailAdd{francesco.sorrenti@unige.ch}
\emailAdd{ruth.durrer@unige.ch}
\emailAdd{martin.kunz@unige.ch}

\abstract{We develop a novel approach to constrain the Hubble parameter $H_0$ and the primordial power spectrum amplitude $A_\mathrm{s}$ using type Ia supernovae (SNIa) data. By considering SNIa as tracers of the peculiar velocity field, we can model their distance and their covariance as a function of cosmological parameters without the need of calibrators like Cepheids; this yields a new independent probe of the large-scale structure based on SNIa data without distance anchors. Crucially, we implement a differentiable pipeline in \texttt{JAX}, including efficient emulators and affine sampling, reducing inference time from years to hours on a single GPU. We first validate our method on mock datasets, demonstrating that we can constrain $H_0$ and $\log 10^{10}A_\mathrm{s}$ within $10\%$ and $15\%$, respectively, using $\mathcal{O}(10^3)$ SNIa. We then test our pipeline with SNIa from an $N$-body simulation, obtaining 6\%-level unbiased constraints on $H_0$ with a moderate noise level. We finally apply our method to Pantheon+ data, constraining $H_0$ at the 15\% level without Cepheids when fixing $A_\mathrm{s}$ to its $\it{Planck}$ value. On the other hand, we obtain 20\%-level constraints on $\log 10^{10}A_\mathrm{s}$ in agreement with $\it{Planck}$ when including Cepheids in the analysis. In light of upcoming observations of low redshift SNIa from the Zwicky Transient Facility and the Vera Rubin Legacy Survey of Space and Time, surveys for which our method will develop its full potential, we make our \texttt{veloce} code publicly available. \href{https://github.com/dpiras/veloce}{\faicon{github}}}


\maketitle

\section{Introduction}
Type Ia supernovae (SNIa) are the best standard candles in cosmology up to date. They have been instrumental in detecting the accelerated expansion of the Universe \citep{SupernovaSearchTeam:1998bnz,SupernovaSearchTeam:1998fmf} and supporting the dark energy hypothesis \citep{SupernovaCosmologyProject:1998vns}. Within the standard flat $\La$CDM cosmology, the recent expansion history is determined by the matter density $\Om_{\rm{m}}$ and Hubble constant $H_0$, which can be both inferred from the luminosity distance-redshift relation $d_\textrm{L}(z)$ in a Friedmann-Lema\^\i tre (FL) universe. For this reason, SNIa are important for joint measurements of $H_0$ and $\Om_{\rm{m}}$, see e.g.\ Ref.~\cite{Brout_2022}. 

However, SNIa are not perfect standard candles, and they need to be normalized with the help of distance anchors; in particular, the Hubble constant is perfectly degenerate with the absolute distance normalization. Moreover, the Universe is not a perfect FL universe, but its matter distribution and geometry are perturbed, while the redshifts of SNIa are affected by the peculiar velocities of their host galaxies. So far, these velocities have mostly been modeled using the velocity field reconstruction approach to convert the CMB-corrected redshifts into so-called `Hubble diagram redshifts', $z_{\rm HD}$ \citep{Carr_2022}, or through a Bayesian hierarchical model marginalizing over unknown cosmological redshifts and peculiar velocities \citep{Tsaprazi25}.

In this work, we consider a different approach to treat velocities, based on Ref.~\citep{Davis11}. We correct the SNIa redshifts for peculiar velocities by adding them to the velocity power spectrum covariance. Like the matter power spectrum, this covariance also depends on the cosmological parameters $H_0$ and $\Om_{\rm{m}}$, in addition to the initial power spectrum described by the amplitude $A_{\rm{s}}$ and the spectral slope $n_{\rm{s}}$. The variance of the SNIa data therefore also contains information about $H_0$ which is completely independent of any distance anchor like the Cepheids. This is particularly interesting in light of the so-called Hubble tension \cite{Efstathiou:2021ocp,Riess:2021jrx,Cosmoverse} as it provides an additional independent way to measure the low-redshift Hubble constant from supernova data.

Other works have looked into constraining the growth of structure from the power spectrum of peculiar velocities, using both simulated and real data. For instance, in Ref.~\cite{macaulay} the authors used both simulated data and various real datasets like the COMPOSITE catalogue~\cite{composite} to estimate the matter power spectrum, finding good agreement with $\Lambda$CDM, while Ref.~\cite{6df_pec} constrained the growth rate of structures using the 6dF Galaxy Survey~\cite{Campbell:2014uia}. Similar analyses were performed in Ref.~\cite{Howlett:2017asq} using the 2MASS Tully–Fisher survey~\cite{Masters_2008}, and in Ref.~\cite{Huterer17} using the Supercal dataset \cite{Scolnic15}, respectively. In Ref.~\cite{Carreres:2023nmf} a forecast of the constraints of the growth rate for ZTF (Zwicky Transient Facility, \cite{Bellm_2019}) was performed using an earlier version of the \texttt{FLIP} package \cite{Ravoux:2025shp}, while in Ref.~\cite{Carreres:2024rji} the impact of peculiar velocities on the measure of $H_0$ for the ZTF second data release~\cite{Rigault:2024kzb} was estimated. Additionally, Ref.~\cite{Hollinger:2025bal} estimated the $H_0$ uncertainty due to peculiar velocities in the Hubble flow. The velocity power spectrum also proved to be useful in testing the standard cosmological model assuming distinct estimates of the matter density, as done in Ref.~\cite{abate_lahav}. 

In our work, we specifically focus on $H_0$ and $A_\textrm{s}$, with a detailed treatment of the velocity covariance. We develop an efficient and differentiable pipeline that combines neural network emulators and affine sampling, and we validate it on mock data and $N$-body simulations. We finally apply our pipeline to Pantheon+ data~\cite{pantheon_light}, showing that in principle supernova surveys can constrain the Hubble parameter without the need for distance anchors. While results from present data are not yet conclusive, we envision that our method will fully unlock its potential with upcoming data like ZTF which include thousands of low-redshift SNIa \cite{Rigault:2024kzb}.

This paper is structured as follows. In Sec.~\ref{sec:model} we describe our theoretical model, while in Sec.~\ref{sec:method} we present the
implementation pipeline. We then validate our approach on a mock dataset, before applying it to an $N$-body simulation as well as to the Pantheon+ data, in Sec.~\ref{sec:results}. We conclude in Sec.~\ref{sec:conclusion}.

\section{Theoretical model\label{sec:model}}

We model the SNIa magnitude data with a Gaussian likelihood, which is given by:
\be \label{eq:likelihood}
\log(\mathcal{L}) = - \frac{1}{2} \biggl[\Delta \mu_{n}\, C^{-1}_{mn}\, \Delta \mu_{m} + \log({\rm det}\,C_{mn})+k\log(2\pi)\biggr],
\ee
where a summation over the repeated indices $m$ and $n$ is implied, and $k$ is the dimension of the covariance $C$, namely the number of SNIa in our dataset. The vector $\Delta\mu_i$ is given by:
\be\label{eq:mu_cases}
\Delta \mu_n =  \begin{cases} \mu_n -\mu_{{\rm ceph},n} + \de \textrm{M} & \text{if} \ n\text{-th SNIa galaxy hosts Cepheids,}\\ 
\mu_n - \mu_\mathrm{model}(z_n^{\rm hel}) + \delta \textrm{M}   & \text{otherwise. }
\end{cases}
\ee 
Here, $\mu_n$ is the observed magnitude of the $n$-th SNIa, and $\delta$M is a normalization parameter, i.e.\ the magnitude shift. 
SNIa in galaxies that also host Cepheids (providing a measurement of their normalized magnitude $\mu_{\rm ceph}$) are uniquely used to determine $\de$M which would otherwise be perfectly degenerate with $H_0$.
The theoretical distance modulus $\mu_{\rm model}$ is related to the luminosity distance via:\footnote{We always indicate the base of the logarithm unless it is the natural one.}
\be \label{eq:mu_model}
\mu_\mathrm{model} = 5 \log_{10} \biggl(  \frac{\langle d_\mathrm{L}(z^{\rm hel}, \bn) \rangle}{\rm Mpc} \biggr) + 25 \, ,
\ee 
where
\bea
d_\mathrm{L}(z,\bn) &=& \bar d_\mathrm{L}(z)\Bigg[1 + \frac{1}{\HH(z) r(z)}\bn\cdot(\bv_\odot-\bv(z,\bn))-\frac{\bn\cdot\bv(z,\bn)}{c}  -  \nonumber  \\
&&  \hspace{-2.1cm}\Phi -\left(1-\frac{1}{\HH r}\right)\left(\!\Psi+ 
\int_{0}^{r}\!\!dr'(\dot\Psi+\dot\Phi)\right)
 +\int_{0}^{r}\frac{dr'}{r} \left(1-\frac{(r-r')}{2r'}\De_\Om\right)(\Psi+\Phi) \Bigg]
\,.\label{e:dLpert1}
\eea
This is is the observable luminosity distance to the measured redshift $z$ at first order in perturbation theory. Here, $\bv_\odot$ is the peculiar velocity of the solar system (measured by the CMB dipole), $\bn$ is the direction of the SNIa, and $\bv(z,\bn)$ its peculiar velocity. Furthermore, $\HH=\HH(z)=\dot{a}/a^2 = H/a$ is the comoving Hubble parameter, $H=H(z)$ is the physical Hubble parameter, $a$ is the scale factor, and $\bar{d_{\mathrm{L}}}$ is the unperturbed luminosity distance defined in Eq.~\eqref{eq:dl} below. Furthermore, $r=r(z)$ is the comoving distance out to redshift $z$ such that $\bar d_L(z) = r(z)(1+z)$, and the variables $\Phi$ and $\Psi$ are the so-called gauge invariant Bardeen potentials that describe scalar perturbations of the metric; in a $\La$CDM cosmology, and at the times and scales relevant here, they are the same and equal to the Newtonian gravitational potential. $\De_\Om$ denotes the 2D Laplacian on the sphere of directions $\bn$. In this full formula the second line has to be evaluated at the position of the supernova given by $z$ and $\bn$ at first order in perturbation theory \cite{Bonvin:2005ps}. In the present work, we only take into account the contributions from velocities given in the first line, which largely dominate at low redshift, $z\ll 1$. At high redshifts $(z\gtrsim 0.5)$ the lensing contribution given by the last integrated term is the most relevant: as it is usually done, we model it as an error proportional to $z^2$. The other relativistic contributions from the gravitational potential are  the integrated Sachs-Wolfe term, Shapiro time delay, and  the gravitational potential at the source. However, they (and some additional terms at the position of the observer which we have not written as they only contribute a constant) remain always very subdominant, see e.g.~Refs.~\cite{Bonvin:2005ps,Pantiri:2024bao} for a numerical evaluation.

The terms we shall consider in this work are the Doppler terms given in the first line of Eq.~\eqref{e:dLpert1}.
We express them in terms of their measured, heliocentric redshift $z^{\rm hel}$ as:
\begin{equation}\label{e:dLpert}
d_\mathrm{L}(z^{\rm hel},\bn) \simeq \bar d_\mathrm{L}(z^{\rm hel})\left[1 + \frac{1}{H(z^{\rm hel})\bar d_\mathrm{L}(z^{\rm hel})}\bn\cdot(\bv_\odot-\bv(z^{\rm hel},\bn))-\frac{\bn\cdot\bv(z^{\rm hel},\bn)}{c} \right] \,.
\end{equation}
This perturbed luminosity distance now depends on direction and can be expanded in spherical harmonics. We shall model it by a monopole, a dipole that models the bulk velocity field, and a quadrupole that models large scale variations of the velocity field. We shall take variations of the velocity field on smaller scales into account in the covariance as a correlated error determined by the velocity power spectrum.

When taking the ensemble average of Eq.~\eqref{e:dLpert} in terms of the CMB-corrected redshift $z= z^\textrm{hel}-(1+z)\bv_\odot\cd\bn/c$ (i.e. correcting $z^{\rm hel}$ for the observer motion, with $c$ ths speed of light),
we find:
\be
\langle d_\mathrm{L}(z,\bn) \rangle = \bar d_\mathrm{L}(z) \, .
\ee
$\bar d_\mathrm{L}(z)$ is the background luminosity distance, which in a flat $\Lambda$CDM model at late times is given by:
\be
\bar d_\mathrm{L}(z) = c(1+z)\!\int_0^z\!\! \frac{dz'}{H(z')}=\frac{c(1+z)}{H_0}\!\int_0^z\! \frac{dz'}{\sqrt{\Omm(1+z')^3+1-\Omm}} = (1+z)r(z)\,. \label{eq:dl}
\ee

Even though the ensemble average of $d_L(z)$ is $\bar d_L(z)$, the measured $d_\mathrm{L}(z)$ does not simply 
average to $\bar d_\mathrm{L}(z)$. This is a manifestation of cosmic variance: what we observe can easily be 1-2$\si$ away from the ensemble average and we have no means to include its error other than by modeling it. Whether one has to do this up to multipole $2$ or $10$ depends on the quality of the data and on the angular coverage of the survey: if instrumental errors are small, more multipoles have to be modeled, while when the angular coverage is narrow, fewer multipoles can be modeled. In the present work we model the monopole, dipole and quadrupole of the Pantheon+ data, which have already been determined and discussed in a previous paper~\cite{Sorrenti:2024a}. These additional contributions are not unexpected and are in good agreement with $\Lambda$CDM, 
as we have shown in Ref.~\cite{Sorrenti:2024a}; they would average out in a true ensemble average over many 
universes, but they do not average out in the data from one universe. Therefore, in our estimator for $\bar d_\mathrm{L}(z)$  
in the case of real supernova data we subtract the maximum-a-posteriori monopole, dipole and quadrupole terms using the \texttt{scoutpip} package.\footnote{\href{https://github.com/fsorrenti/scoutpip}{https://github.com/fsorrenti/scoutpip}}

We write the total covariance matrix $C$ as a sum of an error term, $C^{\rm (e)}_{mn}$ coming from measurement errors in the SNIa magnitudes including a lensing contribution $\propto z^2$, and a velocity-induced term, $C^{\rm (v)}_{ mn}$, that describes the redshift fluctuations due to peculiar velocities:\footnote{For the SNIa in Cepheid-hosting galaxies, the velocity-induced fluctuations affect both SNIa and Cepheids equally, so that this contribution to the covariance is not present for those entries. Moreover, we assume the intrinsic SNIa magnitude dispersion is fixed in $C^{(\mathrm{e})}$, but we find consistent results when allowing it to vary.}
\be \label{eq:full_covariance}
C_{mn}=C^{\rm (e)}_{mn}+C^{\rm (v)}_{mn} \,.
\ee

The error covariance matrix $C^{\rm (e)}$ is the the error covariance of the SNIa data from which the peculiar velocity contribution is subtracted, while the velocity-induced covariance matrix $C^{\rm (v)}$ is due to the difference between the measured redshift $z_n$ and the redshift of the background cosmology $\bar{z}_n$ induced by peculiar velocities. This leads to a difference $\de z_n=(1+z_n)\frac{v_n}{c}$ between the measured and the cosmological redshift, where $v_n=\bn_n\cdot\bv_n$ is the peculiar velocity of the $n$-th object. Here we include this redshift change as an additional fluctuation $\delta\mu_n^{(v)}$ of the distance modulus~\cite{Carreres:2023nmf} given by:
\be \label{eq:velocity_estimator}
\frac{ \log{10}}{5}\delta\mu_n^{(\rm{v})} =\biggl[\frac{c(1+z_n)^2}{d_\textrm{L}(z_n)\,H(z_n)}-1\biggr]\frac{v_n}{c} \, .
\ee
In Appendix~\ref{ap:A} we derive and discuss in detail this result, which is in agreement with Ref.~\cite{Carreres:2023nmf}, 

To obtain the velocity-induced covariance for the magnitudes, we start by considering the covariance of the radial component of the peculiar velocities. In our treatment, we neglect vorticity and first consider the linear velocity power spectrum from scalar perturbations, $P_\textrm{v}$. The covariance matrix of peculiar radial velocities is then of the form:
\be \label{eq:covariance_velocity_velocity_space}
\begin{aligned}
    C^{\rm (v),v}_{mn} &= \int \frac{4\pi k^2 \mathrm{d}k}{(2\pi)^3} W_{mn}(k)\,\langle v_{m}v_{n} \rangle 
    =\int \frac{4\pi k^2 \mathrm{d}k}{(2\pi)^3} W_{mn}(k)\,P_{\rm v}\,(k,z_{m},z_{n})\, ,
\end{aligned}
\ee
where $P_{\rm v}\,(k,z_{ m},z_{ n})$ is the unequal-redshift velocity power spectrum and $W_{mn}$ is a window function determined by the real space positions of the supernovae. To obtain the velocity-induced covariance matrix for the distance modulus $\mu_i$, we multiply Eq.~\eqref{eq:covariance_velocity_velocity_space} with the conversion factor $B_{mn}$ obtained from Eq.~\eqref{eq:velocity_estimator}:
\be \label{eq:conversion_factor}
B_{mn} =\biggl(\frac{5}{c\,\log(10)}\biggr)^2\biggl[\frac{c(1+z_{m})^2}{d_\mathrm{L}(z_{m})\,H(z_{m})}-1\biggr]\biggl[\frac{c(1+z_{n})^2}{d_\mathrm{L}(z_{n})\,H(z_{n})}-1\biggr] \, .
\ee
Note that $B_{mn}$ is independent of $H_0$ as it cancels in the combination $d_\mathrm{L}(z)H(z)$.
The covariance of the SNIa magnitudes from peculiar velocities is then:
\be \label{eq:covariance_velocity}
\begin{aligned}
    C^{\rm (v)}_{mn} =B_{mn}\,& \int \frac{4\pi k^2 \mathrm{d}k}{(2\pi)^3} W_{mn}(k)\,P_{\rm v}\,(k,z_{m},z_{n})
\end{aligned}
\ee
where the window function $W_{mn}$ in Eq.~\eqref{eq:covariance_velocity} is given by:
\be \label{eq:window_gradient}
W_{mn}(k)=\sum_{i,j = 1}^{3}n_{m,i} \, n_{n,j}\int \frac{\diff^2 \hat{k}}{4\pi} \hat{k}_{i} \, \hat{k}_{j} e^{ik \hat{\mathbf k}\cdot(\mathbf r_{m}-\mathbf r_{n})} \, ,
\ee
with ${\mathbf n}_{m} ={\mathbf r}_{m}/r_m $ and ${\mathbf n}_{n}={\mathbf r}_{n}/r_n$ the directions of the $m$-th and $n$-th SNIa respectively, and $i, j$ the spatial indexes. In Appendix~\ref{a:window} we derive an analytic expression for this window function:
\be \label{eq:window_gradient_simple}
\begin{aligned}
W_{mn}(k)=&\frac{1}{3}\cos{\alpha_{mn}} \left[ j_0(kR_{mn}) - 2 j_2(kR_{mn})\right]   +\frac{r_{m}r_{n}}{R_{mn}^2}j_2(kR_{mn})\sin^2{\alpha_{mn}} \, ,
\end{aligned}
\ee
where $\alpha_{ mn}$ is the angle between $\mathbf r_{m}$ and $\mathbf r_{n}$, $j_\ell$ is $\ell$-th order spherical Bessel function and $R_{mn}=|\mathbf{r}_{m}-\mathbf{r}_{n}|$. For $m=n$ we have $W_{mm}=1/3$. These results agree with the more involved derivation given in Ref.~\cite{analytical_window}.

\subsection{The velocity power spectrum}

Considering only linear perturbations, the velocity power spectrum $P_{\rm v}$ is given by:
\be \label{eq:power_spectra_linear_relation}
\begin{aligned}
P_{\rm v}(k, z_{ m},z_{ n})=&\biggl[\frac{\dot D_1(z_m)}{D_1(z_m)(1+z_{m})}\biggr] \biggl[\frac{\dot D_1(z_n)}{D_1(z_n)(1+z_{n})}\biggr] \frac{P_{\delta}(k,z_{ m},z_{ n})}{k^2}\\ =&\biggl[\frac{H(z_{ m})f(z_{ m})}{(1+z_{ m})}\biggr] \biggl[\frac{H(z_{ n})f(z_{ n})}{(1+z_{ n})}\biggr] \frac{P_{\delta}(k,z_{ m},z_{ n})}{k^2} \, ,
\end{aligned}
\ee
where $P_{\delta}(k)$ is the linear matter power spectrum, $D_1$ the linear growth function and $f$ the growth rate. In a flat $\Lambda$CDM cosmology, which we assume in this work, $D_1$ is given by \cite{Euclid:2023qyw}:
\be \label{theory:D1z}
    D_1(z)=\frac{D_0}{(1+z)}\left[ _2F_1 \left ( \frac{1}{3},1;\frac{11}{6};1-\frac{1}{\Om_\text{m}(z)} \right) \right],
\ee
where $_2F_1(a,b;c;d)$ is the confluent hypergeometric function \citep{Abra} and $D_0$ is a normalization constant. The matter density parameter is
\be
\Omm(z)=\frac{\Omm \, (1+z)^{3}}{\Omm \, (1+z)^{3}+(1-\Omm)} \, .
\ee
Furthermore,
\be
\dot D_1(z)= -H(z)(1+z)\frac{dD_1(z)}{dz}\,,
\ee
which is often modeled as $\dot D_1(z) = H(z)f(z)D_1(z)$. Within $\La$CDM, the function $f$ can be determined analytically from Eq.~\eqref{theory:D1z}: it depends only on $\Om_\textrm{m}(z)$ and is well approximated by $f(z)\simeq \Om_\textrm{m}(z)^{0.56}$, see Ref.~\cite{Euclid:2023qyw}. The amplitude of the power spectrum is given by $A_\textrm{s}$ or $\si_8^2$, which is fixed by $A_\textrm{s}$ with a slight dependence on $n_\textrm{s}$. We therefore find that, at linear level:
\be
P_v(k,z_m,z_n)\propto \left\{\begin{array}{cc} 
 f(z_m)f(z_n)H_0^2A_\textrm{s}\,, & \mbox {or}\\  f(z_m)f(z_n)H_0^2\si_8^2 \,. &
 \end{array} \right.
\ee
In the literature it is often assumed that $H_0$ is known, and the amplitude of the velocity power spectrum is used to determine $f\si_8$, see e.g.~Ref.~\cite{Carreres:2023nmf}. Here we
proceed in the opposite way: we fix the model to be $\La$CDM so that $f(z)$ is given and we use the velocity power spectrum to determine $H_0^2A_\textrm{s}$ (noting that we could also replace $A_\textrm{s}$ by $\si_8^2$). We use the full shape of the velocity power spectrum to calculate the velocity covariance matrix; this depends on $H_0$ and $\Om_\textrm{m}$ also via the equality scale.

In order to take into account unequal time correlations of the matter power spectrum, we use the approximation described in Ref.~\cite{zeld}, according to which:
\be \label{eq:zeld_approx}
P_{\delta}(k,z_{ m},z_{ n})=\mathcal{Z}(k,z_{ m},z_{ n})\,P_{\delta}(k,\Bar{z}) \, ,
\ee
with
\be \label{eq:zeld_factor}
\mathcal{Z}(k,z_{ m},z_{ n})=e^{-(k/k_{\rm NL})^2[D_1(z_{ m})-D_1(z_{ n})]^2}\,  
\ee
and the mean redshift $\Tilde{z}$ is defined by
\be \label{eq:z_zeld}
D_1(z_{ m})D_1(z_{ n})=D^2_1(\Tilde{z}) \, .
\ee
The $\mathcal{Z}(k,z_{ m},z_{ n})$ prefactor depends also on the non-linearity scale $k_{\rm NL}$ given by:
\be \label{eq:k_nl}
k_{\rm NL}^{-2}=\frac{1}{12\pi^2}\int_{0}^{\infty}P_{\delta}(k,0)\,\mathrm{d}k \, .
\ee
Using Eq.~\eqref{eq:z_zeld}, we can rewrite Eq.~\eqref{eq:zeld_approx} as:
\be
P_{\delta}(k,z_{ m},z_{ n})=\mathcal{Z}(k,z_{ m},z_{ n})\,D_1(z_{ m})D_1(z_{ n})P_{\delta}(k,0) \, .
 \ee
Combining everything, we obtain the following expression for the velocity covariance:
\be \label{eq:covariance_velocity_final}
\begin{aligned}
    C^{\rm (v)}_{mn} =& 
     \frac{B_{mn}}{2\pi^2} \frac{D_1(z_{ m})~D_1(z_{ n})}{D_1^2(0)}\biggl[\frac{H(z_{ m})f(z_{ m})}{(1+z_{ m})}\biggr]  \biggl[\frac{H(z_{ n})f(z_{ n})}{(1+z_{ n})}\biggr] 
   \\& \int \mathrm{d}k W_{mn}(k)\mathcal{Z}(k,z_{ m},z_{ n})P_{\delta}(k,0) \, .
\end{aligned}
\ee
We compute the linear power spectrum using CAMB~\cite{Lewis:1999bs} up to $k_{\max}=100\ h/$Mpc.

Unlike other works, we do not fix the value of the integral in Eq.~(\ref{eq:covariance_velocity_final}) to the one for the fiducial parameters. Rather, we keep the dependence on cosmological parameters inside the integrand, and tackle the expensive integral for each parameter choice and supernovae pair $(m,n)$ with an efficient pipeline, which we describe in Sec.~\ref{sec:method}. Moreover, in our analysis we have found that the unequal time decoherence factor $\mathcal{Z}(k,z_{ m},z_{ n})$ does not significantly impact the results. This is understood by the fact that supernovae that are at sufficiently different redshifts such that $D_1(z_m)-D_1(z_n)$ is appreciable are so far apart that the power spectrum $P_\de(k)$ is very small for these $k$ values. We note that without $\mathcal{Z}(k,z_{ m},z_{ n})$, $H_0$ and $A_\mathrm{s}$ are nearly degenerate in the linear velocity power spectrum, $P_{\rm{v}}$; the degeneracy is in principle lifted by the equality scale.

\subsubsection{The non-linear velocity power spectrum}
\begin{figure}
  \includegraphics[width=0.5\columnwidth]{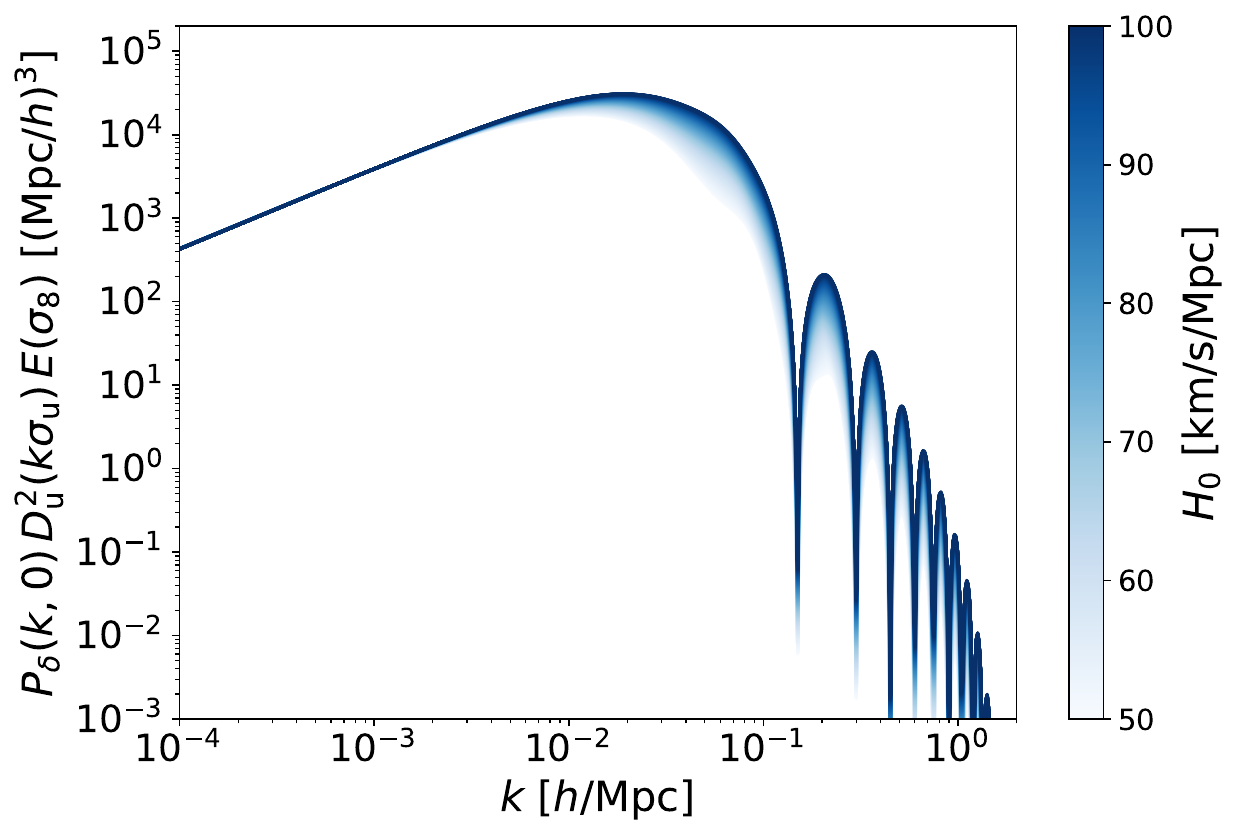}
  \includegraphics[width=0.5\columnwidth]{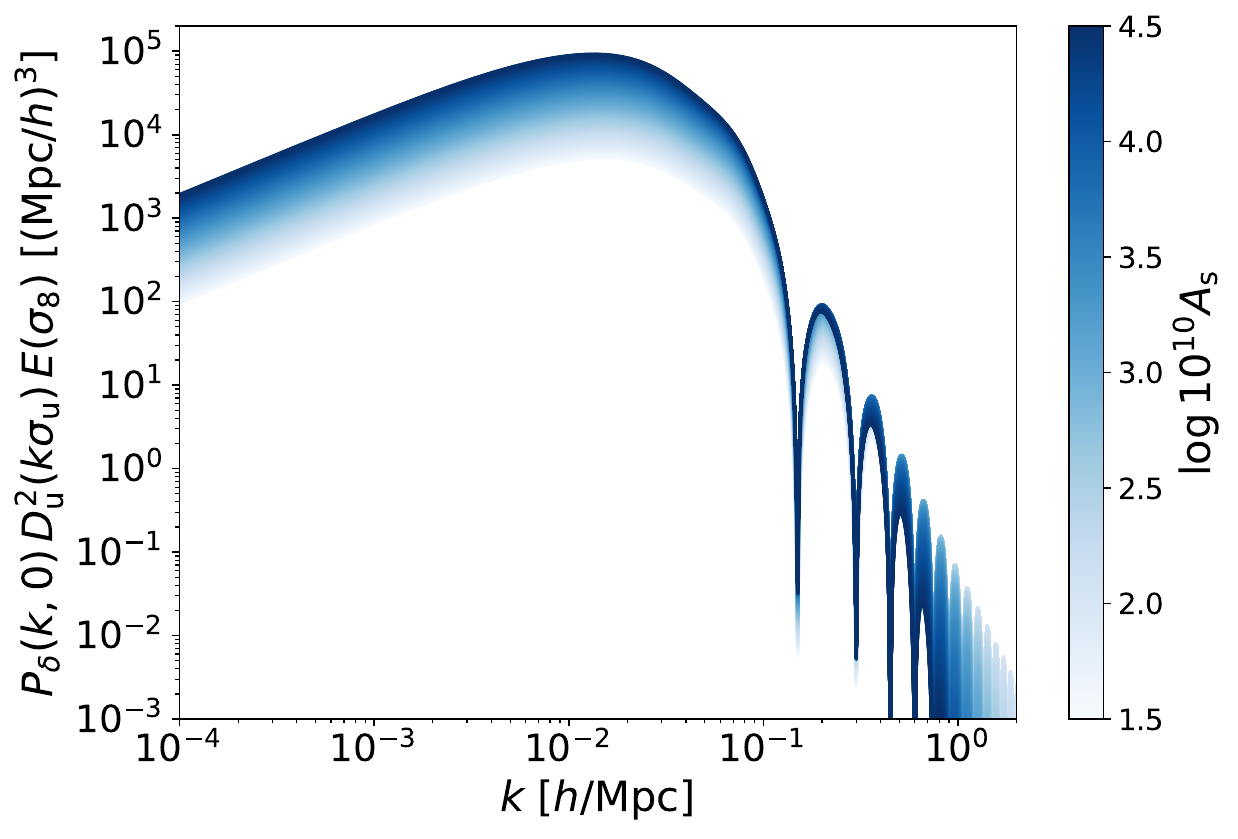}
      \vspace{-0.8cm}
    \caption{Variation of the non-linear power spectrum we assume in this work with $H_0$ (left panel) and $\log_{10}(10^{10}A_\mathrm{s})$ (right panel), with fixed $\log_{10}(10^{10}A_\mathrm{s})=3$ and $H_0=70$, respectively.}
  \label{fig:variations}
\end{figure}

We further consider phenomenological corrections to the linear power spectrum, following Refs.~\cite{koda, Bel:2018awq}. We define the non-linear velocity power spectrum $P^{\rm{NL}}_{\rm v}$ as:
\be
P^{\rm{NL}}_{\rm v}(k, z_{ m},z_{ n})=P_{\rm v}(k, z_{ m},z_{ n})\, D_{\rm u}^2(k\sigma_{\rm u})\,E(\sigma_8) \, ,
\ee 
with
\be 
D_{\rm u}(k\sigma_\textrm{u})={\rm sinc}({k\sigma_\textrm{u}})
\ee
a damping function correction introduced in Ref.~\cite{koda}, and $E(\sigma_8)$ an extra non-linear correction with $3\%$ accuracy for $k<0.7 \ 
h{\rm  Mpc}^{-1}$ based on Ref.~\cite{Bel:2018awq}:
\be 
E(\sigma_8)= e^{-k \max \left[ a_1(\sigma_8)+a_2(\sigma_8)k+a_3(\sigma_8)k^2, 0 \right] } \, ,
\ee
where the $\max$ function has been added to ensure this term is exponentially decaying even for low values of $\sigma_8$. We fix $\sigma_\textrm{u}=13 \ h^{-1}$Mpc, following Ref.~\cite{koda}. The coefficients $a_i$ are linear functions of the root-mean-square of matter density fluctuations $\sigma_8$,
\begin{align}
    a_1(\sigma_8)&=(- 0.817+3.198 \, \sigma_8)\, {\rm M pc}/h \, ,\\
    a_2(\sigma_8)&=(0.877-4.191\, \sigma_8)\, {\rm M pc}^2/h^2\, ,\\
    a_3(\sigma_8)&=(-1.199 + 4.629\, \sigma_8)\, {\rm M pc}^3/h^3\,.
\end{align}
At fixed $n_\textrm{s}$, $\sigma_8$ is proportional to $A_\mathrm{s}$ and given by:
$$\sigma_8=(\sigma_{8,{ \it{Planck}}}\,/A_{\rm s, {\it Planck}}) A_\mathrm{s} \,,
$$
where for $\sigma_{8,{\it Planck}}=0.8102$ and $A_\mathrm{s,\it Planck}=10^{-10}\exp(3.047)=2.105\times 10^{-9}$ we choose the fiducial values from \textit{Planck} (TT,TE,EE+lowE+lensing+BAO)~\cite{Aghanim:2018eyx}. Furthermore, we set $n_{\mathrm{s}}=0.965$ and the baryon density parameter $\omega_\mathrm{b} = 0.02242$. We checked that we obtain consistent results if we assume a higher $\sigma_\textrm{u}=21 \ h^{-1}$Mpc, as in Refs.~\cite{Lai22, Carreres25}, and defer an improved modeling of the velocity power spectrum, following e.g.~\cite{Dam21}, to future work. We show the scaling of the non-linear power spectrum with $H_0$ and $A_\mathrm{s}$ in Fig.~\ref{fig:variations}.

While the linear velocity power spectrum is proportional to $H_0^2A_\mathrm{s}$ and therefore $H_0$ and $A_\mathrm{s}$ are degenerate, this is no longer the case in the non-linear power spectrum as $E(\sigma_8)$ depends on $A_\mathrm{s}$ but not on $H_0$.
However, as we shall see in Sec.~\ref{sec:results}, the presently available data still show a severe degeneracy between the value of $H_0$ inferred from the velocity covariance and $A_\mathrm{s}$.

\subsection{Velocity dispersion}
\label{sec:dispersion}
The velocity power spectrum gives in principle the non-linear peculiar velocity field as determined with $N$-body simulations. However, measuring this field with a finite, relatively small number of tracers adds shot noise that can be taken into account as an additional velocity dispersion~\cite{pryor,godwin}. We model this in the form:
\be
C_{mn}^{\rm(v),disp} = B_{mn}\sigma_{\rm disp}^2\de_{mn} \, ,
\ee
only for those galaxies not hosting a Cepheid. This additional dispersion has also been interpreted as velocities due to virial motions of galaxies in clusters, see Ref.~\cite{koda}. The factor $B_{mn}$ again converts the velocity variance into a magnitude variance.
We leave the value of $\sigma_{\rm disp}$ free in our analysis.

\section{Methodology}
\label{sec:method}

We develop a differentiable Markov chain Monte Carlo (MCMC) pipeline combining efficient emulators and an affine sampler, running on a single graphics processing unit (GPU) and converging $10^5$ times faster than traditional approaches. The emulators' ranges are reported in Table~\ref{table:priors}, which also includes the prior ranges for the MCMC analyses. 

\subsection{Pipeline description}

The integral in Eq.~\eqref{eq:covariance_velocity_final} (with $P_\mathrm{v}$ replaced by the non-linear velocity 
power spectrum, $P_\mathrm{v}^\mathrm{NL}$) is computationally expensive: assuming a dataset containing $10^3$ objects, we need to compute it for 
all $\sim 5\cdot10^5$ unique pairs of objects in the dataset at each likelihood evaluation. For each choice 
of cosmological parameters, i.e.\ for each step of the MCMC, this requires more than half million evaluations of the integral, with the number increasing quadratically with the number of objects in the survey. For this reason, we develop an emulator to evaluate the integral, computing each element of the covariance matrix using a neural network based on the \texttt{CosmoPower} emulation framework \citep{SpurioMancini22}. 

We first randomly select one million combinations of redshift pairs and angles from the Pantheon+ dataset, and concatenate them with cosmological parameters sampled uniformly from the ranges indicated in Table~\ref{table:priors}, which also act as the priors for the inference step. We compute the value of the integral using the \texttt{FFTLog} algorithm to account for the Bessel functions \citep{TALMAN197835, Fang:2019xat},\footnote{\href{https://github.com/xfangcosmo/FFTLog-and-beyond}{https://github.com/xfangcosmo/FFTLog-and-beyond}} separating between the case $\alpha_{mn}=0$ (which does not involve Bessel integrals) and $\alpha_{mn} \neq 0$. We then train two emulators (one for $\alpha_{mn}=0$ and one for $\alpha_{mn} \neq 0$) to predict the value of the integral given the values of $z_1$, $z_2$, $\alpha_{12}$, $\Omega_{\rm{m}}$, $H_0$ and $\log\left( 10^{10} A_{\rm{s}}\right )$, using a mean absolute error loss function to be robust to outliers. We find that each emulator converges in a few hours on a single GPU with a median performance at the sub-percent level on held-out test data, so we are confident that its predictions are accurate. We note that our emulators are trained on redshifts and angles drawn from the Pantheon+ dataset, so in principle one would need to train again in case a new dataset is considered; however, we tested that the performance was still satisfactory even on the $N$-body simulation data we consider in this paper. In future work, we will explore a more general emulator trained on a uniform distribution of redshifts and angles, which should be applicable to a wider variety of datasets. 

\begin{table}
\centering
\caption{Prior distributions of the parameters for the training of the emulator (first three columns), and for inference (all columns). Uniform distributions are indicated with $\mathcal{U}$. The units of $H_0$ are km/s/Mpc, while the units of $\sigma_{\textrm{disp}}$ are km/s. }
\begin{tabular}{ c c c c c c }
\toprule
 \textbf{Parameter} & $\Omega_{\rm{m}}$ & $H_0$ & $\log\left( 10^{10} A_{\rm{s}}\right )$ & $\log_{10}(\sigma_\textrm{disp}^2)$ &  $\delta $M \\
 \midrule 
 \textbf{Prior range} & $\mathcal{U}[0.1, 0.5]$ & $\mathcal{U}[50, 100]$ & $\mathcal{U}[1.5, 4.5]$ & $\mathcal{U}[2, 6]$ & $\mathcal{U}[-0.5, 0.5]$ \\
 \bottomrule
 \end{tabular}
 \label{table:priors}
 \end{table}
 
We write the likelihood in Eq.~\eqref{eq:likelihood} using {\tt JAX} \citep{Jax18repo, Jax18}, a library that enables differentiable and efficient {\tt{Python}} code optimized for GPUs. To sample the posterior distribution, we use the {\tt JAX} version of {\tt affine},\footnote{\href{https://github.com/justinalsing/affine}{https://github.com/justinalsing/affine}} a parallelized affine-invariant sampler based on {\tt emcee} \cite{emcee, autocorr}; our {\tt CosmoPower} emulator is integrated into the likelihood through {\tt CosmoPower-JAX} \cite{Piras:2023aub}.\footnote{\href{https://github.com/dpiras/cosmopower-jax}{https://github.com/dpiras/cosmopower-jax}} The combination of a parallelized sampler and a neural network emulator, both running on a GPU, significantly accelerates our analysis: we sample the posterior distribution for a single model using 22 walkers (18 in the $N$-body simulation case) and 5000 steps in no more than one hour using a single A100 80GB GPU after training the emulator, which is done once and requires a single day including the generation of the training set. We extrapolate that a similar analysis employing 32 CPUs and no emulator would require $\sim 25$ years, since each likelihood call requires the evaluation of the integral in Eq.\ \eqref{eq:covariance_velocity_final} for each pair of supernovae. For our $\sim 10^6$ pairs this takes about $2$ days. This means that the use of our differentiable pipeline provides a total speed-up greater than $10^5$. We make our emulator and likelihood codes publicly available.\footnote{\href{https://github.com/dpiras/veloce}{https://github.com/dpiras/veloce}}

The MCMC convergence is achieved when the number of steps $N > 50\tau_{\rm max}$, with $\tau_{\rm max}$ the largest integrated autocorrelation time $\tau$ among all the parameters (see Ref.~\cite{autocorr} for more details). After convergence, we discard $2 \, \lfloor \tau_{\rm max} \rfloor$ steps from the chain as burn-in \citep{burnin}. The contour plots are obtained using \texttt{chainconsumer v0.34} \citep{Hinton2016}. We also verified our MCMC results against a profile likelihood approach \citep{Planck:2013nga}, finding consistent results.

\section{Results}\label{sec:results}

\begin{figure}
  \includegraphics[width=0.5\columnwidth]{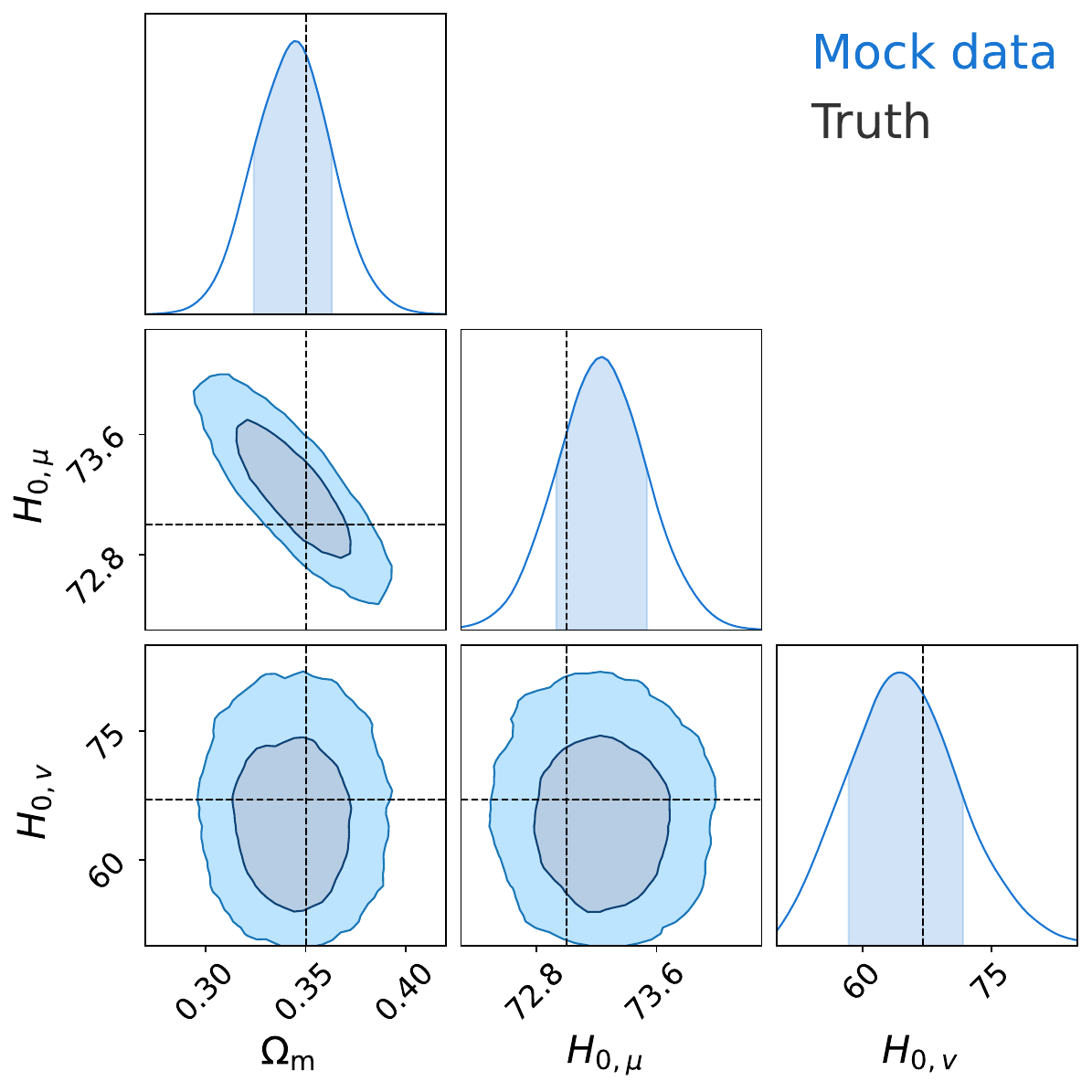}
  \includegraphics[width=0.5\columnwidth]{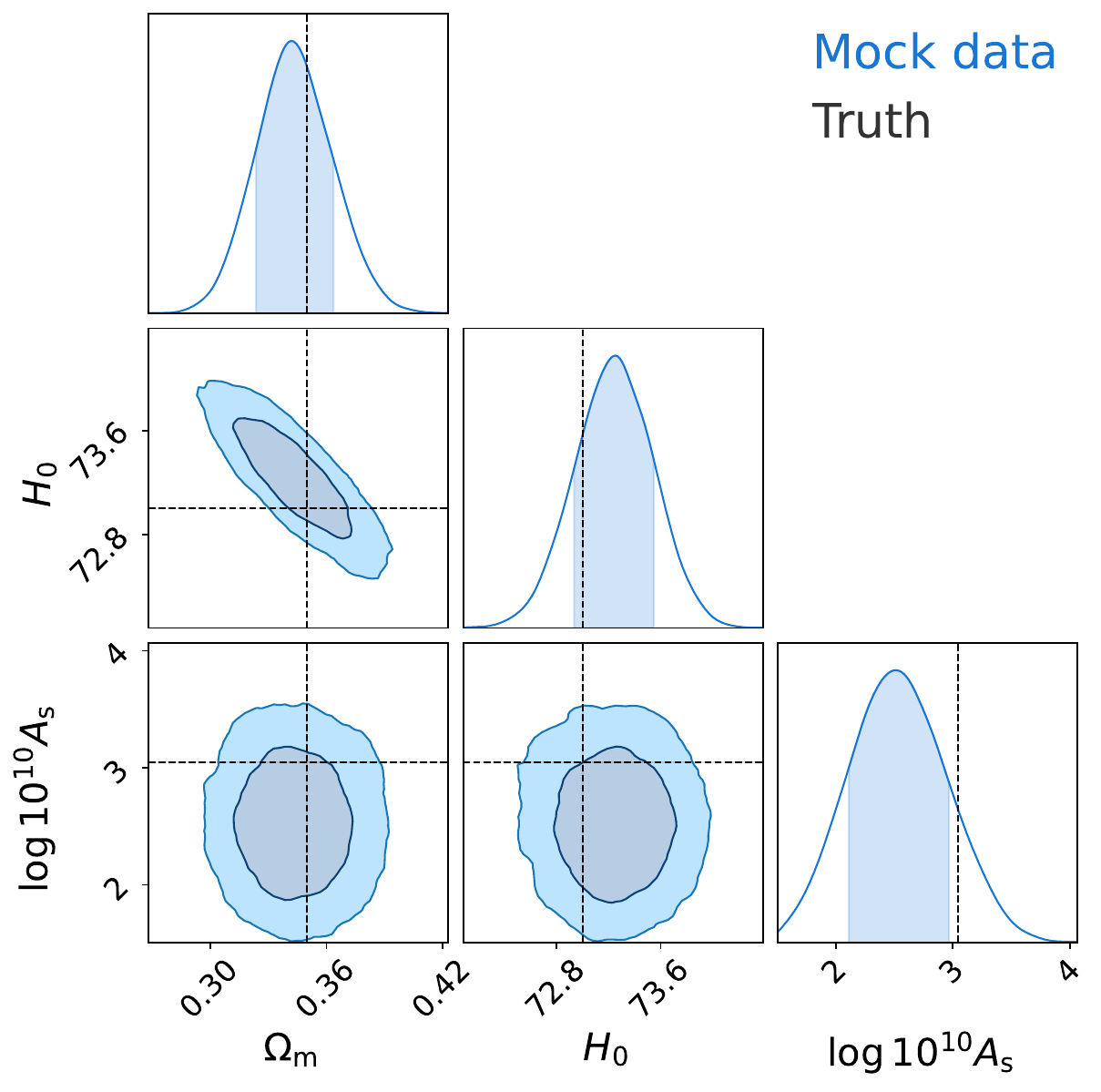}
      \vspace{-0.8cm}
    \caption{Posterior distribution for the mock test, with the bands indicating the 68\% and 95\% credible intervals. In the left panel we choose $H_{0,\mu}=73$ km/s/Mpc for the distances and $H_{0,\rm{v}}=67$~km/s/Mpc for the velocity power spectrum; additionally, $\Om_{\textrm{m}}=0.35$ and $\log_{10}(10^{10}A_\mathrm{s})=3.047$. For the analysis in the right panel we assume $H_{0,\mu}=H_{0,\rm{v}}$, but vary $A_\mathrm{s}$ which is degenerate with $H_{0,\rm{v}}$. We do not include Cepheids in these mock datasets, and simply set $\de \textrm{M}\equiv 0$ which leads to an artificially small error in $H_{0,\mu}$. In all figures throughout the paper, $H_0$ values are in units of km/s/Mpc.}
  \label{fig:mock_6773_doubleH0}
\end{figure}

\subsection{Mock dataset}
We generate a mock dataset following the steps described in Appendix~\ref{ap:mock}. This dataset is modeled with redshifts affected by velocities generated from the velocity covariance matrix in Eq.~\eqref{eq:covariance_velocity_velocity_space}, and distance moduli with errors given by $C^{\rm (e)}$ as in Pantheon+. In order to study how well we can recover $H_0$ from the peculiar velocities and from the distance moduli, we choose different input values of $H_0$ for the velocity covariance, denoted $H_{0,{\rm \rm{v}}}=67$ km/s/Mpc, and the distance modulus, denoted $H_{0,\mu}=73$ km/s/Mpc. As a first step with this mock data, we do not add a velocity dispersion as described in Sec.~\ref{sec:dispersion}, which is present not only in real data but also in $N$-body simulations.

When fixing $A_\mathrm{s}$, we can recover both $H_{0,\rm{v}}$ and $H_{0,\mu}$ within $1\sigma$, as we show in the left panel of Fig.~\ref{fig:mock_6773_doubleH0}. In the right-hand side plot of Fig.~\ref{fig:mock_6773_doubleH0}, using the same dataset, we assume just one physical value of $H_0 = H_{0,\mu}= H_{0,\rm{v}}$: since the velocity covariance was generated with a different value of $H_0$ this reproduces now a smaller value of $A_\mathrm{s}$ with an only very little increase in goodness of fit ($\log\mathcal{L} =743$ vs $740$). As expected within linear perturbation theory, the result mainly depends on $H^2_0A_\mathrm{s}$ and a 10\% increase in $H_0$ can be compensated with an about 20\% decrease in $A_\mathrm{s}$. 
We also see that mapping the redshift error onto the magnitude error does not induce a bias in the analysis of the mock dataset, although for larger datasets a more involved analysis may be necessary~\cite{March:2011xa}.

\subsection{$N$-body simulation}

As a more realistic test dataset, we use a subset of the simulated distances created by the relativistic $N$-body code {\tt gevolution} \cite{Adamek:2016zes} to study the impact of cosmic structure on the Hubble diagram in Ref.~\cite{Adamek:2018rru}. For this, a relativistic ray-tracing through the simulation volume has been performed. The simulation covers a cosmological volume of $(2.4 \ {\rm Gpc}/h)^3$, the metric is sampled on a regular grid of $7680^3$ points and the matter density is followed by $7680^3$ mass elements. The simulation assumes a flat $\Lambda$CDM cosmology with $H_0=67.556$~km/s/Mpc, $\Omega_\textrm{m} = 0.312046$ and $A_\mathrm{s} = 2.215 \times 10^{-9}$, corresponding to $\log10^{10}A_\mathrm{s}=3.098$. The lightcone was recorded on a circular pencil beam covering $450 \ {\rm deg}^2$. More details can be found in Ref.~\cite{Adamek:2018rru}.

\begin{figure}
  \includegraphics[width=0.5\columnwidth]{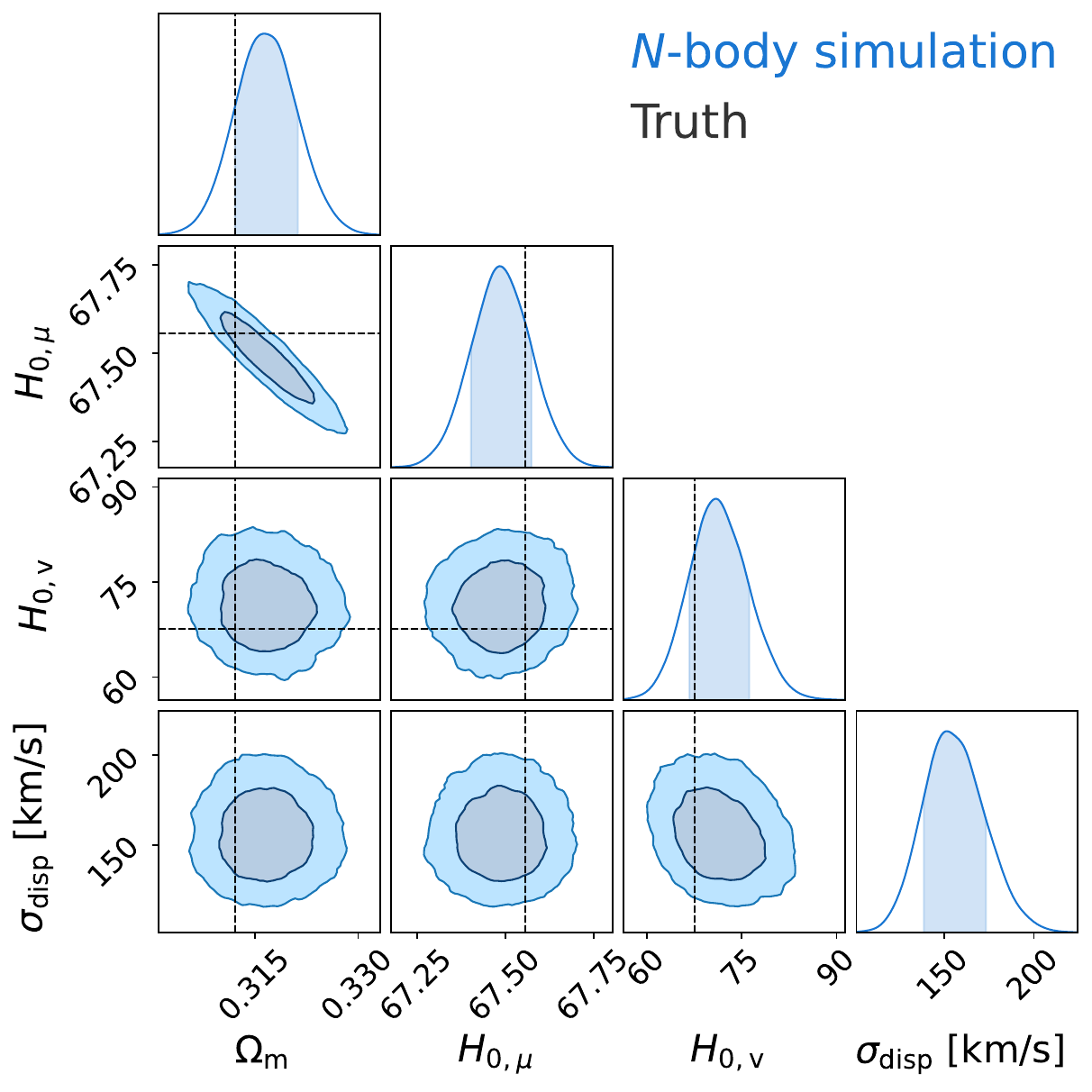}
  \includegraphics[width=0.5\columnwidth]{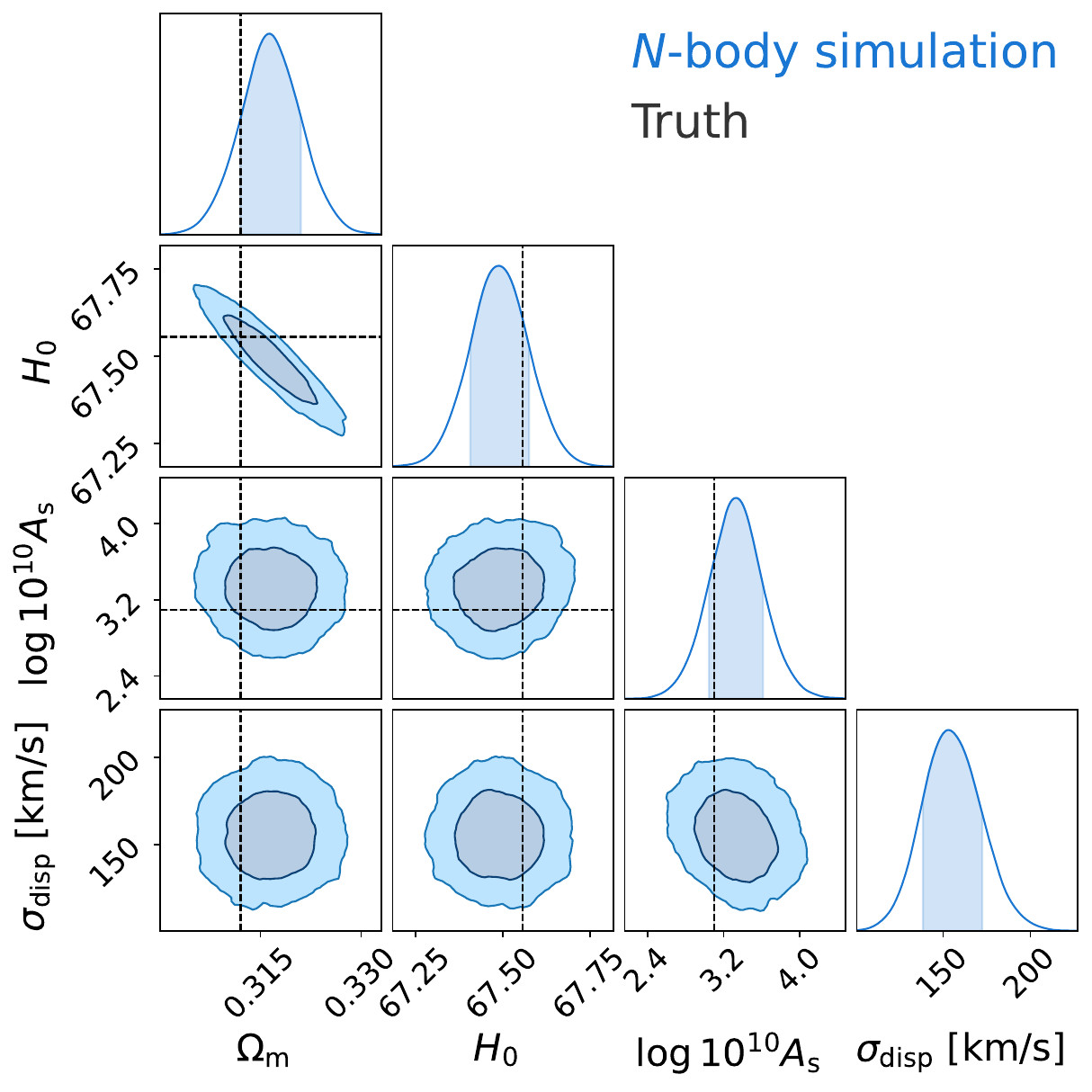}
  \vspace{-0.8cm}
    \caption{Results from the $N$-body simulation run with \texttt{gevolution}. The blue contours indicate the 68\% and 95\% credible intervals, while the black dashed lines represent the input cosmological parameters for the $N$-body simulation. The left panel shows that, given $A_\textrm{s}$, we can determine $H_0$ either via the distance moduli, $H_{0,\mu}$ or via the peculiar velocity covariance, $H_{0,\rm{v}}$. In the right panel, we consider only one value of $H_0=H_{0,\mu}=H_{0,\rm{v}}$, and we use the velocity covariance to determine $A_\textrm{s}$.}
  \label{fig:gevolution_results}
\end{figure}

\begin{figure}
\centering  \includegraphics[width=0.9\columnwidth]{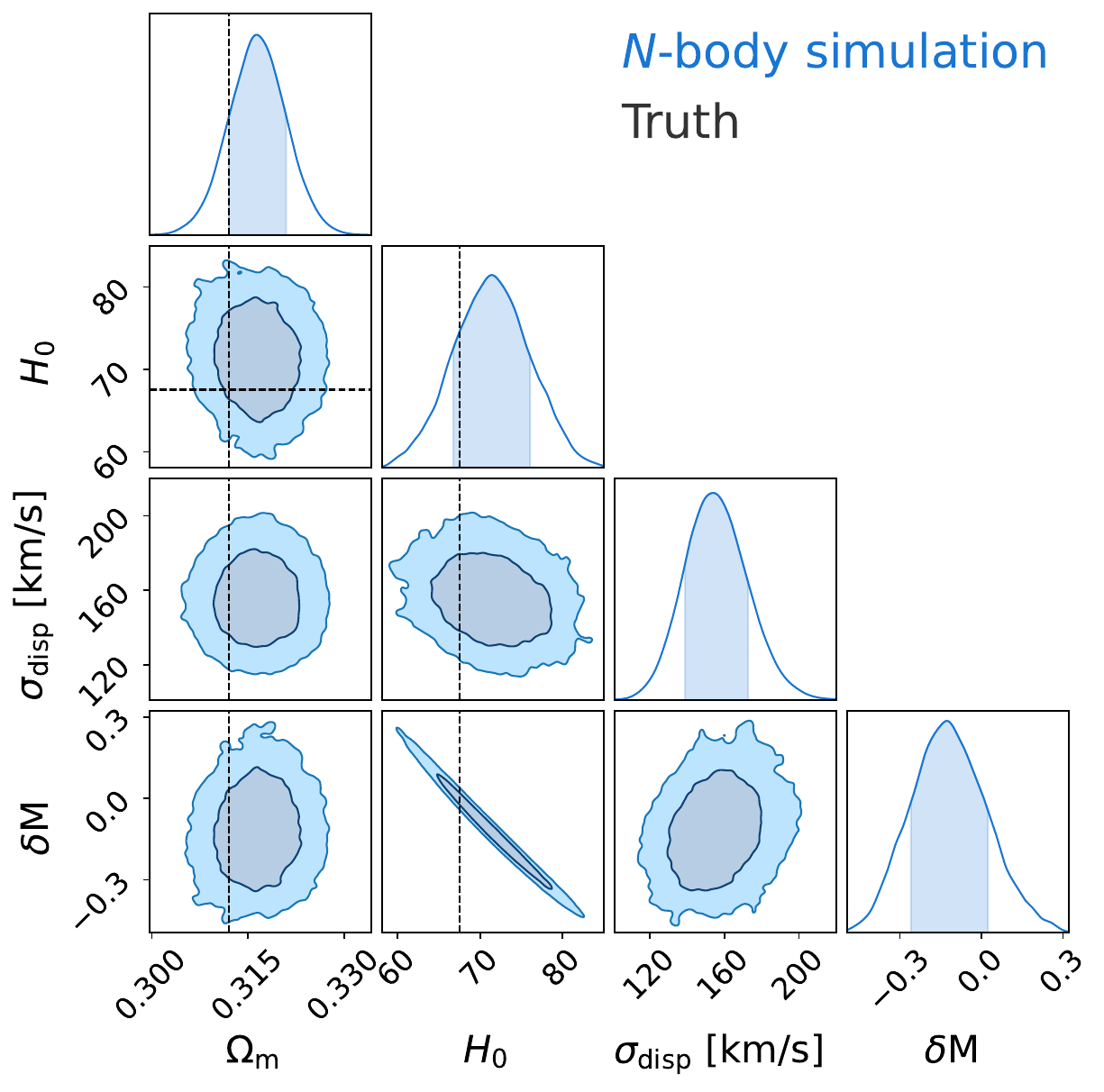}
    \vspace{-0.3cm}
    \caption{Results from the $N$-body simulation \texttt{gevolution} `without cepheids', namely with $\delta$M acting as calibration parameter. The blue contours indicate the 68\% and 95\% credible intervals, while the input cosmological parameters for the $N$-body simulation are shown as dashed lines. Fixing $A_\textrm{s}$ to the input value, we use the velocity covariance to determine $H_0$ without the help of anchors. The degeneracy with $\de$M, while not being completely lifted, is significantly reduced.
    }
  \label{fig:gevolution_noceph}
\end{figure}

These $N$-body distances contain the velocity and lensing contributions by construction, but have essentially vanishing intrinsic scatter. To make them more realistic, we add random noise to the data with a given constant standard deviation $\sigma_0=0.03$, roughly corresponding to one order of magnitude less than Pantheon+, to test a scenario with smaller error bars corresponding to a higher number of SNIa.\footnote{Assuming the true intrinsic spread is 0.1 mag but uncorrelated for different supernovae (in different galaxies), then multiplying the observed number of supernovae by a factor 10 would correspond roughly to a reduction of the statistical part of the standard deviation by a factor $\sqrt{10}\simeq 3$; this translates into an intrinsic magnitude reduction from 0.1 mag to about 0.03 mag.} We then write the error covariance matrix as:
\[
C^{\rm (e)}_{mn} = \left( \sigma_0^2 + \sigma_l^2 z^2 \right) \delta_{mn}
\]
for the chosen value of $\sigma_0$ and for a lensing contribution of $\sigma_l = 0.055$ \cite{Jonsson:2010wx}. To this we add the velocity covariance $C^{\rm (v)}_{mn}$ and the velocity dispersion $C_{mn}^{\rm(v),disp}$ as described in Sec.~\ref{sec:model}.

From the large number of distances to halos available in this data set, we randomly select 1701 halos in the redshift range $\left[0.01, 1\right]$ with a redshift distribution similar to the one of the Pantheon+ data. Given the box size and resolution, we expect that the large-scale structure is correctly resolved down to significantly smaller redshifts. We also have checked that the monopole, dipole and quadrupole contributions to the luminosity distance are negligible in this near `pencil beam' data.

We perform the same analyses with the $N$-body simulations as in the next section for the real supernova data. We first fix $A_\textrm{s}$ and assume two different values of $H_0$, one for the distance modulus, $H_{0,\mu}$, and one for the peculiar velocities, $H_{0,{\rm v}}$. The results, shown in the left panel of Fig.~\ref{fig:gevolution_results}, indicate that we manage to recover $\Omega_\textrm{m}$ and both $H_0$ values within 1$\sigma$, with a velocity dispersion of $\sigma_{\rm{disp}}= 155^{+18}_{-16}$ km/s. In the right panel of Fig.~\ref{fig:gevolution_results} we further show that, assuming a single $H_0$, we also retrieve the correct value of $\log10^{10}A_\textrm{s}$, with a consistent value of $\sigma_{\rm{disp}}$.

In Fig.~\ref{fig:gevolution_noceph} we show the results of a parameter estimation mimicking an analysis without any Cepheid anchors. Clearly, the degeneracy between $H_0$ and the distance shift $\de$M is considerable, but it is partially broken by the velocity covariance, allowing us to recover well the input value of $H_0$, for which we find $H_0 = 71.4^{+4.7}_{-4.6} \ \rm{km/s/Mpc} $.
The input $\Om_\textrm{m}$ is also well recovered, while the required velocity dispersion of $\sigma_{\rm disp} = 155^{+18}_{-16}$ km/s is in good agreement with the Pantheon+ data results discussed below. While the non-linear corrections in the velocity covariance matrix can in principle break the degeneracy between $A_\textrm{s}$ and $H_0$, we still find in this analysis a strong degeneracy, namely we can either determine $H_0$ or $A_\textrm{s}$, but not both at the same time. This might be due to the fact that there are very few nearby supernovae in this dataset, so that the non-linear corrections to the velocity covariance are not very important here. We will however always show results including the non-linear modeling of the velocity covariance.

\subsection{Pantheon+ data}
\label{sec:pantheon}
\begin{figure}
  \includegraphics[width=0.5\columnwidth]{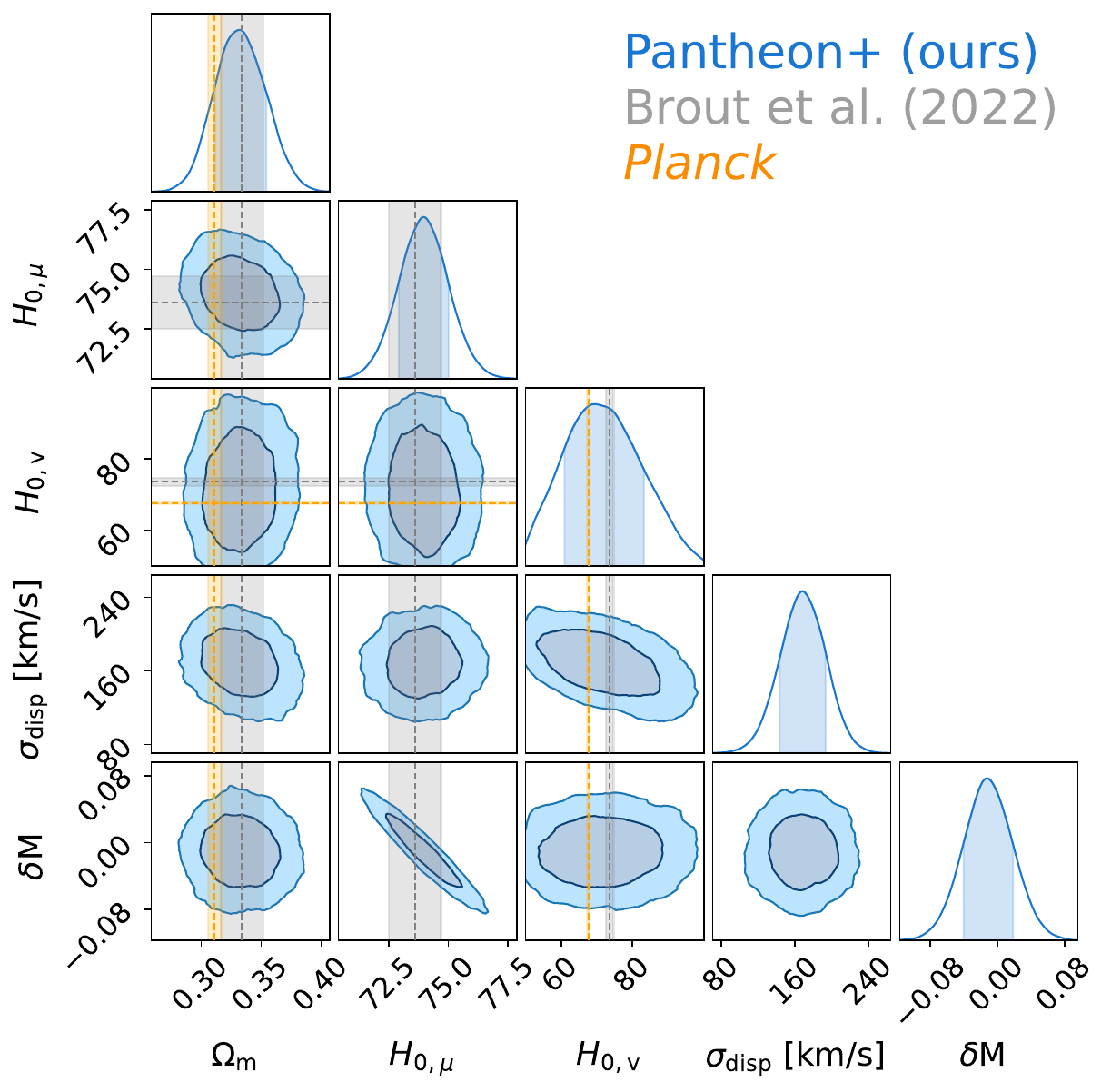}
  \includegraphics[width=0.5\columnwidth]{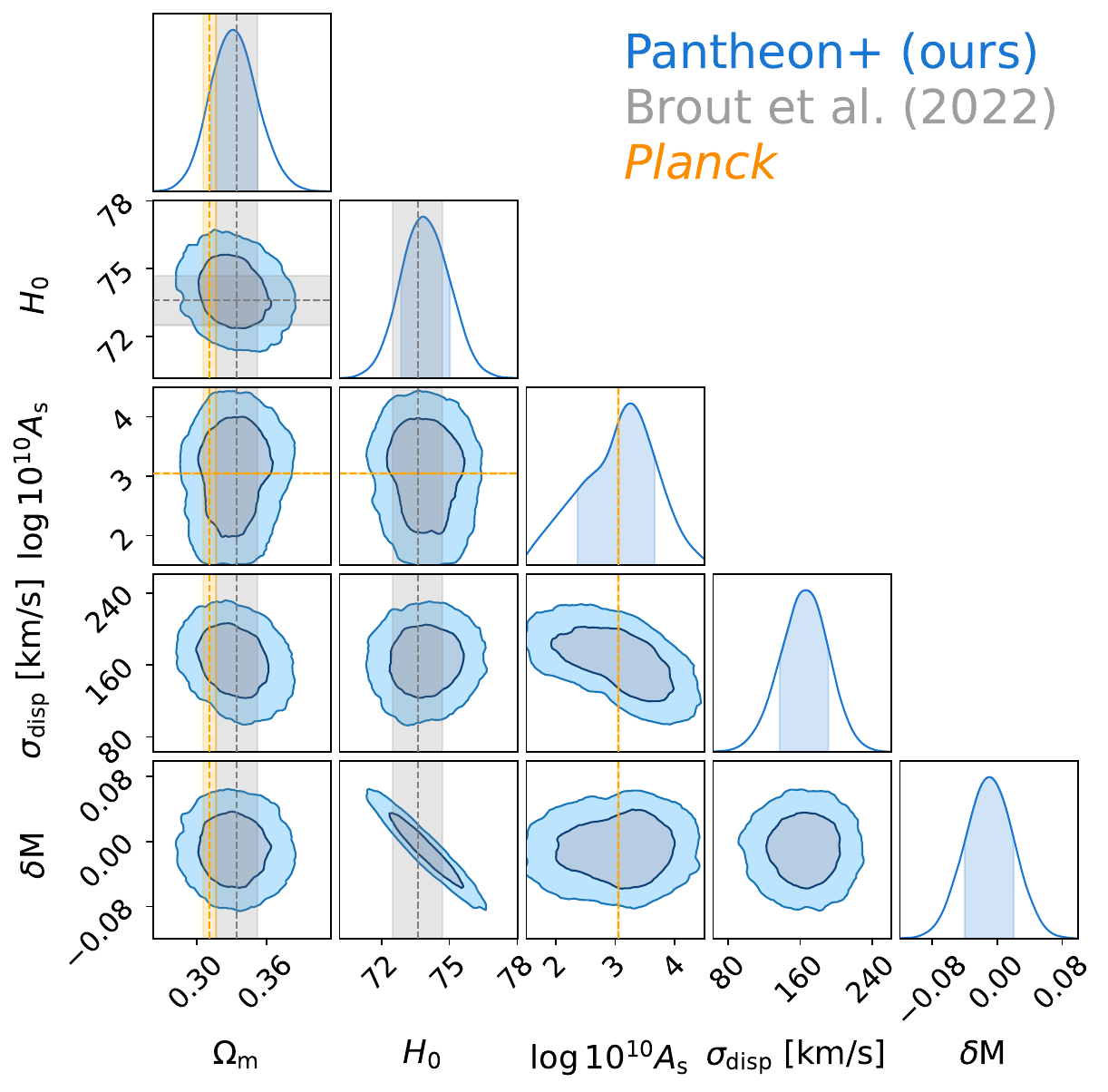}
    \vspace{-0.8cm}
    \caption{68\% and 95\% posterior contours from the Pantheon+ data including Cepheid information. In the left panel, we model the peculiar velocities with fixed $(A_\textrm{s},~n_\textrm{s})$ from the \textit{Planck} data, and fit $H_0$ from the distance moduli ($H_{0,\mu}$) and from the velocity covariance ($H_{0,\rm{v}}$) independently. The inferred value for $H_{0,\rm{v}}$ has large error bars and cannot distinguish between the value inferred from the traditional supernova analysis (gray bands) and the \textit{Planck} value (orange bands). In the right panel, we model the data with one value for $H_0$ and use the velocity covariance to determine $A_\textrm{s}$. Again, the errors are still rather large with present data, but demonstrate the potential of our approach.}
  \label{fig:pantheon_results}
\end{figure}

\begin{figure}
\centering  \includegraphics[width=0.8\columnwidth]{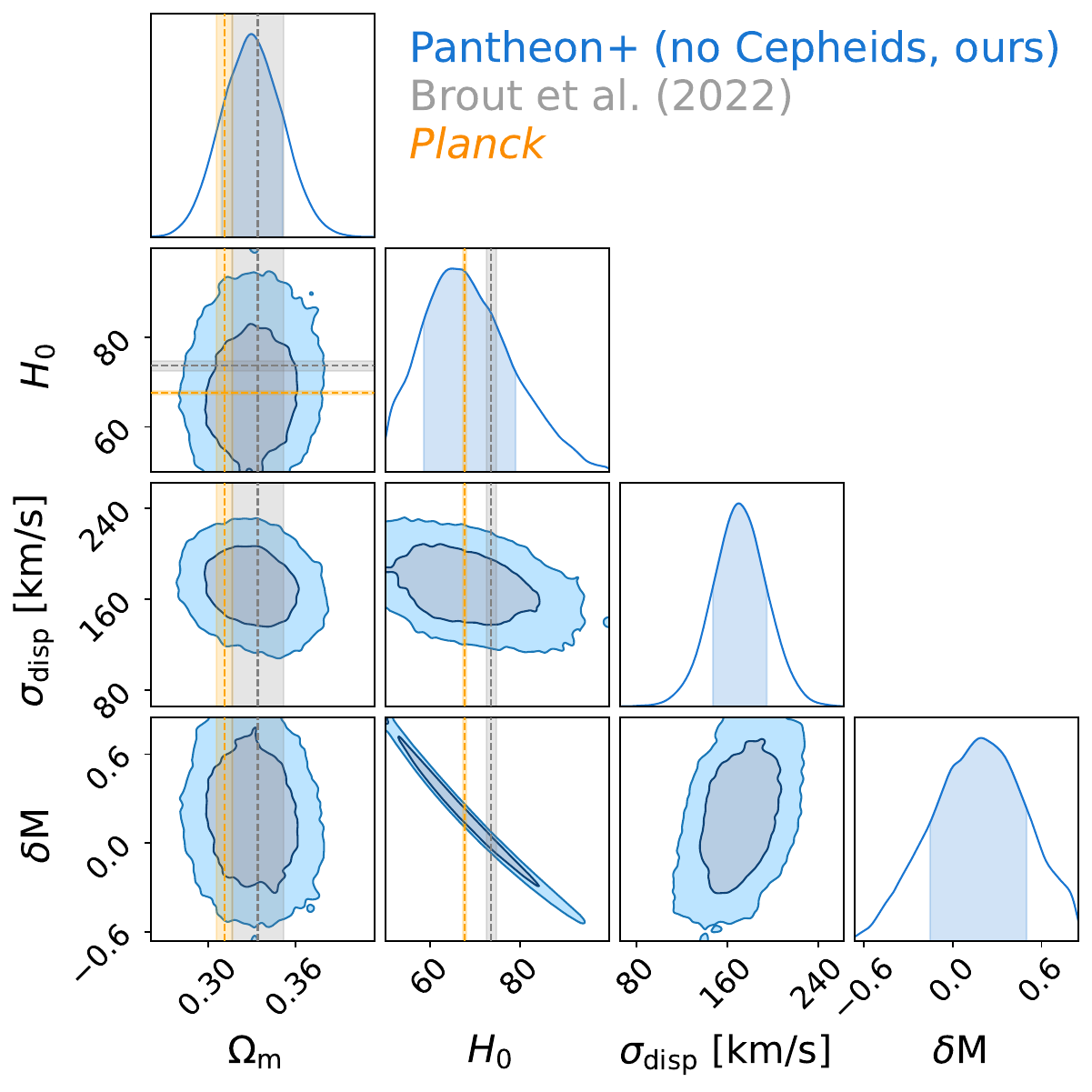}
      \vspace{-0.5cm}
    \caption{68\% and 95\% posterior contours from the Pantheon+ analysis without using Cepheid anchors. The degeneracy between $\de $M and $H_0$ is broken only by our treatment of the velocity covariance. The \textit{Planck} values for $H_0$ and $\Om_\textrm{m}$ are indicated as orange bands, while in gray we report the values from the latest Pantheon+ analysis \cite{Brout_2022}.}
  \label{fig:pantheon_results_noceph}
\end{figure}

As in our previous analyses (see Refs.~\cite{Sorrenti:2022zat, Sorrenti:2024a, Sorrenti:2024ugq}), we use the Pantheon+SH0ES dataset providing 1701 lightcurves \cite{Brout_2022, pantheon_light}; of these, 77 SNIa are in galaxies hosting Cepheids, whose absolute distance modulus $\mu_\mathrm{ceph}$ is known.
For the analysis of the Pantheon+ data, we start from the Pantheon+ covariance matrix with the statistical and systematic contributions from peculiar velocities subtracted, and we add our peculiar velocity covariance matrix $C^{\text{(v)}}_{mn}$. When fixing $A_\textrm{s}$ and $n_\textrm{s}$ to the \textit{Planck} values, the velocity covariance matrix depends on $H_0$ and $\Omega_\textrm{m}$ via the power spectrum $P_\textrm{v}$. We first fix $A_\mathrm{s}$ to the \textit{Planck} value and allow for an independent $H_0$ for both the distance modulus, $H_{0,\mu}$ and the velocity covariance, $H_{0,\rm{v}}$; the result is shown in the left panel of Fig.~\ref{fig:pantheon_results}. Clearly, $H_{0,\rm{v}}$ has very wide contours and cannot distinguish between the values preferred by the distance Pantheon+ data \cite{Brout_2022} (gray bands) and the \textit{Planck} data \cite{Aghanim:2018eyx} (orange bands). On the other hand, we can use the distance moduli and the velocity covariance to jointly model the value of $H_0$, $\Om_\textrm{m}$ and $A_\textrm{s}$, which we show in the right panel of Fig.~\ref{fig:pantheon_results}. In this way, supernovae provide an independent value of the scalar perturbation amplitude $A_\textrm{s}$ which is nearly independent of \textit{Planck} data (only the value $n_\textrm{s}$ is adopted from \textit{Planck}).

We finally remove the calibrators from the Pantheon+ dataset and perform an analysis assuming a single $H_0$ while fixing $A_\mathrm{s}$ to its \textit{Planck} value. In this instance only, we extend the $\de $M prior range to $\left[ -1, 1\right]$. The results, presented in Fig.~\ref{fig:pantheon_results_noceph}. cannot yet distinguish between the \textit{Planck} value and the value from the standard SNIa analysis; however, we start breaking the degeneracy between $H_0$ and $\delta$M, finding:
\[
H_0 = 67.7^{+11.3}_{-9.2} \ {\rm km/s/Mpc} \,.
\]
All numerical results for our Pantheon+ analyses are presented in Table~\ref{tab:results}.

\begin{table}[ht]
    \setlength\tabcolsep{4 pt} 
\centering
     \caption{Parameter constraints (median $\pm$ 68\% credible intervals) for the Pantheon+ analyses, with all details available in Sec.~\ref{sec:pantheon}. All analyses find $\Omm=0.33\pm0.02$. We have $H_{0, \mu}=H_0$ in the second and third rows.\vspace{0.1cm}} 
    \begin{tabular}{cccccccc}
    \toprule
         &  $H_{0,\mu}$&$H_{0,\rm{v}}$& $\log 10^{10}A_\mathrm{s}$ & $\sigma_{\rm disp}$& $\delta$M\\         Analysis&\scriptsize{[km/s/Mpc]}&\scriptsize{[km/s/Mpc]}& &\scriptsize{[km/s]} &\\
		\midrule 
With Cepheids, fixed $A_\mathrm{s}$ (Fig.~\ref{fig:pantheon_results})& $74.0^{+1.0}_{-1.0} $  & $72.0^{+12.0}_{-11.0}$ &-  & $169^{+25}_{-25}$& $-0.01^{+0.03}_{-0.03}$   \\
With Cepheids, free $A_\mathrm{s}$ (Fig.~\ref{fig:pantheon_results}) & $73.9^{+1.1}_{-1.1} $ & -& $3.1^{+0.5}_{-0.8}$ & $165^{+26}_{-28}$ & $-0.01^{+0.03}_{-0.03}$  \\
No Cepheids, fixed $A_\mathrm{s}$ (Fig.~\ref{fig:pantheon_results_noceph}) & $67.7^{+11.3}_{-9.2} $ & - & -& $171^{+24}_{-24}$ & $0.18^{+0.32}_{-0.34}$   \\
    \bottomrule
    \end{tabular}
\label{tab:results}
\end{table}

\section{Conclusions}
\label{sec:conclusion}
In this paper, we introduce a novel method to use SNIa peculiar velocities to constrain cosmological parameters without the need for anchors such as the Cepheids, by exploiting the $H_0$ dependence of the covariance matrix of the peculiar velocities. So far, our analysis constrains either $A_\mathrm{s}$ or the Hubble parameter in the velocity power spectrum, $H_{0,\rm{v}}$, since the linear velocity power spectrum only depends on the product $H_0^2A_\mathrm{s}$. With more low redshift supernovae, we shall also become more sensitive to non-linear corrections which break this degeneracy.
We developed an efficient differentiable pipeline and validated our approach on mock datasets and on $N$-body simulations. We then applied it to the Pantheon+ data, finding that $H_{0,\rm{v}}=72.0^{+12.0}_{-11.0}$ km/s/Mpc for the double $H_0$ analysis, while for the single $H_0$ analysis without Cepheids we find $H_0=67.7^{+11.3}_{-9.2}$ km/s/Mpc.

Our work showcases a new path to address tensions between early and late Universe probes \citep{Verde:2019ivm}. Presently, there is a 
considerable debate concerning the use of anchors for supernova measurements~\cite{Freedman:2024eph,Riess:2024vfa,Freedman:2025nfr}; our method is independent of these 
anchors, depending only on the peculiar velocity power spectrum. While present data does not have sufficiently many low redshift supernova to allow a precise 
measurement of the velocity power spectrum, the proposed method will achieve its full potential with upcoming surveys like ZTF \cite{Bellm_2019} and the Vera Rubin Legacy Survey of Space and Time (LSST, \cite{lsst}), providing significantly more data especially at low redshift. As peculiar velocities 
contribute most to the Hubble diagram at low redshifts, especially in the analysis of ZTF with an order of 
magnitude more supernovae with redshifts $z<0.1$, our method promises to unlock precise cosmological constraints without relying on anchors.

We plan to extend our analysis in several ways. For instance, we will implement a differentiable version of the low multipoles of the luminosity distance, based on Ref.~\cite{Sorrenti:2024a}, which will allow us to jointly sample the multipoles and the cosmological parameters.
We will also implement other contributions to the peculiar velocities, such as the vorticity power 
spectrum. Vorticity is usually neglected when modeling velocities; nevertheless, some works (e.g.~\cite{Jelic-Cizmek:2018gdp}) have shown that it is an important contribution to the velocity power spectrum at late time, exactly when peculiar velocities become more and more relevant in the Hubble diagram. We 
will also further evaluate the robustness of our new implementations with more refined $N$-body simulations. 

Most importantly, we plan to apply our routine to larger supernova datasets as the LSST and, in particular, the ZTF. The contribution from ZTF will be particularly relevant since it is expected to observe many supernovae at low redshift; for example, the upcoming Data Release 2.5 is anticipated to contain thousands of objects at $z\leq 0.3$~\cite{Rigault:2024kzb}.
In order to obtain competitive constraints which can distinguish between the presently advocated values of $H_0$, we will need about 10 times more SNIa at low redshift ($z\lesssim 0.1$) than the present dataset; Pantheon+ contains 664 SNIa with $z<0.1$, excluding the 77 SNIa in galaxies hosting Cepheids. In our upcoming analyses, we will expand our pipeline to jointly fit the covariance matrix along with the parameters of the Tripp formula \citep{Tripp98}, as well as to include selection and systematic effects such as Malmquist bias and chromatic intrinsic scatter.
Our approach can also be modified to work with other distance relations like the Tully-Fisher relation~\cite{Tully:1977fu} and the Fundamental Plane relation~\cite{Djorgovski:1987vx, dressler}, as well as to be applied to datasets like CosmicFlows-4~\cite{Tully:2022rbj}, containing distances to 55877 galaxies collected into 38065 groups.

\section{Data availability}
The Pantheon+ dataset and distance modulus covariances are available in the official repository \url{https://github.com/PantheonPlusSH0ES/DataRelease}. The data and the code to reproduce our analysis, together with the emulator for the velocity covariance power spectrum, are publicly available at this GitHub repository: \href{https://github.com/dpiras/veloce}{https://github.com/dpiras/veloce} \href{https://github.com/dpiras/veloce}{\faicon{github}}.

\acknowledgments
We thank Anthony Carr, Richard Watkins, Pedro Ferreira and Alex Kim for useful discussions; in particular, we would like to thank Anthony Carr for sharing the Pantheon+ covariance without the peculiar velocity contribution. DP, FS and MK acknowledge financial support from the Swiss National Science Foundation. DP was additionally supported by the SNF Sinergia grant CRSII5-193826 “AstroSignals: A New Window on the Universe, with the New Generation of Large Radio-Astronomy Facilities”. The computations underlying this work were performed on the Baobab cluster at the University of Geneva. This work used data originally generated by a grant from the Swiss National Supercomputing Centre (CSCS) under project ID s710.


\appendix

\section{Derivation of $\de z$}\label{ap:A}

We assume that the shift in $\bar{z} = z + \delta z_i$ due to the peculiar velocity of supernova $i$ is small, so that we can write:
\be
\mu(\bar{z}_i) = \mu(z_i) + \mu'(z_i) \delta z_i\,, \quad \delta z_i =\frac{v_i/c}{1+z_i} \, ,
\ee
where $v_i=\bv(\bn_i,z_i)\cd\bn_i$ is the radial peculiar velocity of supernova $i$,\footnote{We neglect corrections in the redshift due to the gravitational field which are usually much smaller.} and $\bn_i$ its position.
The derivative of the distance modulus $\mu =5\log_{10}(d_\mathrm{L}/1{\rm Mpc}) + 25$ is given by
\be
\mu'(z_i) = \frac{\partial \mu}{\partial z_i} = \frac{5}{\log(10)} \frac{d_\mathrm{L}'(z_i)}{d_\mathrm{L}(z_i)} \, .
\ee
We can invert the relation to obtain the shift in $z$:
\be
\delta z_i = \frac{\mu(\bar{z_i})-\mu(z_i)}{\mu'(z_i)}
=  \frac{\log(10)}{5} \frac{d_\mathrm{L}(z_i)}{d_\mathrm{L}'(z_i)} \left[ \mu(\bar{z_i})-\mu(z_i) \right] \, .
\ee

From the definition of the background $d_\mathrm{L}(z)$ we have that:
\be
\frac{d_\mathrm{L}'}{d_\mathrm{L}} = \frac{1}{1+z} + \frac{1+z}{H(z) d_\mathrm{L}(z)}
= \frac{1}{1+z} \left( 1 + \frac{(1+z)^2}{H(z) d_\mathrm{L}(z)} \right) \, .
\ee
Furthermore, also $d_\mathrm{L}(\bar z)$ is proportional to the source frequency squared, adding a term $-2\de z/(1+z) d_\mathrm{L}$ which inverts the sign of the first term in the final expression (see Ref.~\cite{Bonvin:2005ps} for a detailed derivation including also the gravitational terms).
Using the peculiar velocity term of Ref.~\cite{Bonvin:2005ps} in the equation for $\delta z$ and writing $\delta \mu^{(v)} = \mu(\bar{z})-\mu(z)$, we find:
\bea
\delta z_i &=& \frac{\log(10)}{5} \delta \mu^{(v)}_i \frac{1+z_i}{\frac{c(1+z_i)^2}{H(z_i) d_\mathrm{L}(z_i)}-1} \, , \eea
or equivalently
\bea
v_i &=& \frac{\log(10)}{5} \delta \mu^{(v)}_i \frac{c}{\frac{c(1+z_i)^2}{H(z_i) d_\mathrm{L}(z_i)}-1} \,,
\eea
which agrees with Eq.\ \eqref{eq:velocity_estimator}.

\section{Derivation of the window function}\label{a:window}
To calculate the window function as in Eq.~\eqref{eq:window_gradient}, we must determine the integrals:
\bea \label{ea:window_gradient}
W_{mn}(k) &=& \sum_{i,j=1}^{3}\bn_{m,i}\bn_{ n,j}\int \frac{\diff\Om_{\hat{k}}}{4\pi} \hat{\bk}_{i} \, \hat{\bk}_{j} e^{ik \hat{\mathbf k}\cdot(\mathbf r_{m}-\mathbf r_{n})}\,= \sum_{i j}\bn_{m,i}\bn_{n,j}W_{mnij}(k) \, ,
\eea
where ${\mathbf n}_{m} ={\mathbf r}_{m}/r_m $ and ${\mathbf n}_{n}={\mathbf r}_{n}/r_n$ are the directions of the $m$-th and $n$-th SNIa respectively. To simplify the calculation we orient the coordinate system in $\bk$-space such that $\mathbf r_{m}-\mathbf r_{n}$ is in the $z$-direction, and both $\mathbf r_{m}$ and $\mathbf r_{ n}$ are in the $\varphi=0$ plane; hence $r_{ m 2}=r_{n 2}=0$. Furthermore, it is easy to see that the $\varphi$-integral of $W_{mnij}(k)$ vanishes whenever $i\neq j$. Using $\cos\theta =\mu$ such that $\diff\Om_{\hat{k}} =\sin\theta d\theta d\varphi = d\mu d\varphi$ and setting $|{\mathbf r_{ m}-\mathbf r_{n}}| =R_{mn}$, we obtain:
\bea
W_{mn33}(k) &=& \frac{1}{2}\int_{-1}^1\mu^2 e^{ikR_{mn}\mu}
\nonumber\\
&=&
\frac{2kR_{mn}\cos(kR_{mn})+((kR_{mn})^2-2)\sin(kR_{mn})}{(kR_{mn})^3} \, , \\
W_{mn11}(k) &=& \frac{1}{4\pi}\int_0^{2\pi}\cos^2\varphi\int_{-1}^1(1-\mu^2) e^{ikR_{mn}\mu}
\nonumber\\
&=&
\frac {-2kR_{mn}\cos(kR_{mn})+2\sin(kR_{mn})}{(kR_{mn})^3} \, .
\eea
We now use that: 
$$W_{mn}(k)= (n_{m3}n_{n3}+n_{m1}n_{n1})W_{mn33} + n_{m1}n_{n1}(W_{mn11}-W_{mn33})\,.$$
Furthermore, one easily verifies that:
$$
W_{mn33}(k)=\frac 13 (j_0(kR_{mn}) - 2 j_2(kR_{mn}))
$$
and
$$
W_{mn11}(k)-W_{mn33}(k)= j_2(kR_{mn}) \, .
$$
We use also that:
$$n_{m3}n_{n3}+n_{m1}n_{m1} =\cos\al_{mn} \, ,$$
and the elementary triangle relation
$$
\frac{\sin\theta_m}{r_n} = \frac{\sin\theta_n}{r_m} = \frac{\sin\al_{mn}}{R_{mn}}\,,
$$
which implies
$$
n_{m1} = \sin\theta_m = r_n\frac{\sin\al_{mn}}{R_{mn}}
$$
and
$$
n_{n1} = \sin\theta_n = r_m\frac{\sin\al_{mn}}{R_{mn}} \,.
$$
Putting it all together we arrive at the final result:
\be \label{ea:window_gradient_simple}
\begin{aligned}
W_{mn}(k)=&\frac{1}{3}\cos{\alpha_{mn}} (j_0(kR_{mn}) - 2 j_2(kR_{mn})) + \frac{r_{ m}r_{ n}}{R_{mn}^2}j_2(kR_{mn})\sin^2{\alpha_{mn}}\,.
\end{aligned}
\ee

\section{Mock dataset}\label{ap:mock}
For the sake of simplicity and to avoid numerical instabilities, we create our mock dataset starting from a subset of Pantheon+ without Cepheids containing 1457 individual lightcurves. These are obtained by iteratively removing the element corresponding to the largest absolute component of the eigenvector associated with the most negative eigenvalue, until the resulting velocity covariance matrix is semi-positive definite. We correspondingly subsample the error covariance $C^{\rm (e)}_{mn}$ obtained from the full Pantheon+ covariance. This procedure is necessary due to the presence of objects with similar, or even identical, sky position in the Pantheon+ collection, which make the Cholesky decomposition in Eq.~(\ref{eq:cholesky}) unfeasible.
 
We choose the heliocentric redshifts of the original dataset as background redshift $\bar z$ of the mock dataset. We then define the mock observed redshifts $z^{\rm mock}_i$ for the $i$-th object as:

\be
z^{\rm mock}_i = \bar{z}_i + \delta z_i \, ,
\ee
with the peculiar velocity contribution

\be
\delta z_i = \sum_j D_{ij} g_j \, .
\ee
Here, $g_j$ is sampled from a Gaussian distribution with mean 0 and variance $1$, while $D_{ij}$ is obtained from the Cholesky decomposition of our velocity covariance:
\be
\label{eq:cholesky}
\frac{C^{\rm (v),z}}{c^2}= D D^{\rm T},
\ee
with $c$ being the speed of light and 
\be
C^{\rm (v),z}=C^{\rm (v),v}(1+z_m)(1+z_n). 
\ee
$C^{\rm (v),v}$ is the covariance in Eq.~\eqref{eq:covariance_velocity_velocity_space} obtained assuming $H_{0,\rm{v}}=67$ km/s/Mpc, $\Om_{\textrm{m}}=0.35$ and $\log_{10}(10^{10}A_\mathrm{s})=3.04$.
Finally, we define the mock distance modulus as:

\be
\mu_i^{\rm mock}=\mu_i(\bar{z}_i)+h_i\,\sigma^{\mu}_i
\ee
with $h_i$ sampled from a Gaussian distribution with mean 0 and variance 1, and $(\sigma^{\mu}_i)^2$ is the $i$-th element of the diagonal of the Pantheon+ covariance without peculiar velocity correction. $\mu_i(\bar{z}_i)$ is obtained from Eq.~\eqref{eq:mu_model} assuming the same cosmology as for $z^{\rm mock}_i$ except for $H_{0,\mu}=73$ km/s/Mpc.


\bibliographystyle{JHEP}

\bibliography{refs} 


\end{document}